\documentclass[showpacs,10pt,twocolumn,prl]{revtex4-1}
\usepackage{amsmath}
\usepackage{amssymb}
\usepackage{graphics}
\usepackage{epsfig}
\usepackage{CJK}
\usepackage{color}

\begin{document}

\title{Nearly-room-temperature ferromagnetism and tunable anomalous Hall effect in atomically thin Fe$_{4}$CoGeTe$_{2}$}
\author{Shaohua Yan,$^{\dag, a,b}$ Hui-Hui He,$^{\dag, a,b}$ Yang Fu,$^{\dag, a,b}$ Ning-Ning Zhao,$^{a,b}$ Shangjie Tian,$^{c,a,b}$ Qiangwei Yin,$^{a,b}$ Fanyu Meng,$^{a,b}$ Xinyu Cao,$^{d}$ Le Wang,$^{a,b}$ Shanshan Chen,$^{a,b}$  Ki-Hoon Son,$^{e}$ Jun Woo Choi,$^{e}$ Hyejin Ryu,$^{e}$ Shouguo Wang,$^{c}$ Xiao Zhang,$^{*, d}$ Kai Liu,$^{*, a,b}$ and Hechang Lei$^{*, a,b}$}

\affiliation{$^{a}$Department of Physics and Beijing Key Laboratory of Optoelectronic Functional Materials \& MicroNano Devices, Renmin University of China, Beijing 100872, China.\\
		$^{b}$Key Laboratory of Quantum State Construction and Manipulation (Ministry of Education), Renmin University of China, Beijing 100872, China.\\
		$^{c}$School of Materials Science and Engineering, Anhui University, Hefei 230601, China.\\
		$^{d}$State Key Laboratory of Information Photonics and Optical Communications \& School of Science, Beijing University of Posts and Telecommunications, Beijing 100876, China.\\
		$^{e}$Center for Spintronics, Korea Institute of Science and Technology (KIST), Seoul 02792, Korea.}
\date{\today}
	
\begin{abstract}

Itinerant ferromagnetism at room temperature is a key ingredient for spin transport and manipulation.
Here, we report the realization of nearly-room-temperature itinerant ferromagnetism in Co doped Fe$_{5}$GeTe$_{2}$ thin flakes. The ferromagnetic transition temperature $T_{\rm C}$ ($\sim$ 323 K - 337 K) is almost unchanged when thickness is down to 12 nm and is still about 284 K at 2 nm (bilayer thickness).
Theoretical calculations further indicate that the ferromagnetism persists in monolayer Fe$_{4}$CoGeTe$_{2}$. 
In addition to the robust ferromagnetism down to the ultrathin limit, Fe$_{4}$CoGeTe$_{2}$ exhibits an unusual temperature- and thickness-dependent intrinsic anomalous Hall effect.
We propose that it could be ascribed to the dependence of band structure on thickness that changes the Berry curvature near the Fermi energy level subtly.
The nearly-room-temperature ferromagnetism and tunable anomalous Hall effect in atomically thin Fe$_{4}$CoGeTe$_{2}$ provide opportunities to understand the exotic transport properties of two-dimensional van der Waals magnetic materials and explore their potential applications in spintronics.

\end{abstract}

\maketitle


\section{Introduction}

Since the discovery of graphene\cite{Novoselov1}, two-dimensional van der Waals (2D vdW) materials have been studied intensively because of their extraordinary physical properties such as valley polarization, Ising superconductivity and so on\cite{Mak,Saito}.
Moreover, the wide variety of physical properties together with the easy combination of different vdW materials due to the weak interlayer bonding further provide great opportunities to realize the multi-functional devices and discover novel emergent phenomena in 2D vdW materials and their heterostructures\cite{Geim,Novoselov2,LiuY}.
Among the unique properties of 2D vdW materials, the long-range ferromagnetism at 2D limit is long-sought because unlike that in three-dimensional (3D) systems, the long-range ferromagnetism in isotropic 2D system is fragile against thermal fluctuation according to the Mermin-Wagner theorem \cite{Mermin}, which impedes the realization of ferromagnetism in 2D systems at finite temperature. 
The breakthroughs in 2D ferromagnetism were achieved in thin layers of Cr$_{2}$Ge$_{2}$Te$_{6}$\cite{Gong1} and CrI$_{3}$\cite{Bevin1} with Curie temperature $T_{\rm C}<$ 100 K and in these systems the anisotropy of spin interaction stabilizes the long-range ferromagnetism at finite temperature\cite{Gibertini}.
Due to their atomic-scale thickness with tunable magnetic properties by electric field, chemical doping, strain field etc., as well as the feasibility of rapid fabrication of various 2D heterostructures with stacking-dependent properties,  
2D vdW magnets opens up a wide range of possibilities for basic science and device applications, such as the studies of critical behaviours of 2D magnetism, the electrical manipulation and detection of spin, spintronics at 2D limit with reduced size of device\cite{Gibertini}.

In order to utilize the 2D ferromagnets in the practical applications, the improvement of magnetic transition temperature up to room temperature is an utterly important task.
Although many efforts have been made to explore vdW ferromagnetic (FM) materials with high $T_{\rm C}$, such as  MnSe$_{x}$\cite{Dante}, VSe$_{2} $\cite{Bonilla,Wang2}, CrSe\cite{Zhang},  Cr$_{x}$Te$_{y}$\cite{SunXD,Wang1,ZhangXQ,Wen}, Fe$_{x}$GeTe$_{2}$ ($x=$ 3, 4, 5)\cite{May,May3,Andrew,Deng,FeiZY,Seo,Tan,LiZY}, Fe$_{3}$GaTe$_{2}$\cite{ZhangGJ} etc, 
the atomically thin 2D vdW materials with intrinsic nearly-room-temperature ferromagnetism are still scarce and some results are still under debate\cite{Bonilla,Wang2,Deng,Daniel}. 
Moreover, when compared to the 2D vdW ferromagnetic materials grown by epitaxy whose properties are sensitive to growth conditions, raw materials, substrate effects etc., the study of 2D magnets obtained by mechanical exfoliation is important to understand the intrinsic properties of these materials.
Among the known vdW ferromagnets, Fe$_{x}$GeTe$_{2}$ family are promising material systems because not only can the $T_{\rm C}$s of bulk materials be increased significantly by varying the iron content or element doping\cite{Seo,ChenX,Stahl,Tian,Andrew2,May2}, but the $T_{\rm C}$ of Fe$_{x}$GeTe$_{2}$ thin flakes could also be enhanced above 300 K by ionic gating or exerting pressure\cite{Deng,LiZY}. 
For example, the $T_{\rm C}$  of four-layer Fe$_{3}$GeTe$_{2}$ (F3GT) rises from around 150 K to more than 300 K with ionic gating\cite{Deng} and the $T_{\rm C}$ of Fe$_{5}$GeTe$_{2}$ (F5GT) thin flakes is enhanced to more than 400 K under pressure\cite{LiZY}.  
More importantly,  very recently the quantum magnetic imaging technique reveals that the Co doped F5GT thin flake with thickness $t$ about 16 nm exhibits intrinsic ferromagnetism above room temperature ($T_{\rm C}\sim$ 310 K) at ambient pressure without ionic gating\cite{ChenH}. Moreover, Monte Carlo simulations predict that the monolayer (ML) of this material still has a nearly-room-temperature $T_{\rm C}$ ($\sim$ 270 K)\cite{ChenH}. 

In this work, we study the ferromagnetism of mechanically exfoliated Fe$_{4}$CoGeTe$_{2}$ (F4CGT) thin flakes. 
From the measurements of anomalous Hall resistivity, it is found that the $T_{\rm C}$ of bulk material ($\sim$ 320 K) is almost intact when the thickness $t>$ 12 nm and then decreases mildly for thinner samples, but the $T_{\rm C}$ above 280 K still can be retained in the bilayer (BL) F4CGT ($t\sim$ 2 nm). Theoretical calculations indicate that the ML F4CGT is still a ferromagnet with easy-plane ($ab$ plane) magnetic anisotropy.
Further analysis shows that the intrinsic anomalous Hall conductivity (AHC) exhibits non-monotonic dependence on temperature $T$ and $t$. It may be related to the evolution of Berry curvature with the position of Fermi level $E_{\rm F}$ and band structure, both of which changes with $T$ and $t$.

\section{Experimental Section}

\noindent\textbf{Single crystal growth and structural characterization.} 
Single crystals of F4CGT were grown by chemical vapor transport method\cite{Chen}. Powders of Fe (98 \% purity), Co (99.5 \% purity), Ge (99.999 \%) and Te (99 \%) in a molar ratio of 4 : 1 : 1 : 2  were put into a quartz tube with 80 mg I$_{2}$ flakes (99.999 \%). The tube was evacuated and sealed at 0.01 Pa. The sealed quartz ampoule was placed vertically in box furnace and the temperature was heated to 1033 K for 7 hours. After that, the temperature was held there for five days and the natural temperature gradient in the furnace was used to grow single crystals, which is about 20 K from bottom to top of ampoule. Shiny plate-like crystals with lateral dimensions of up to several millimeters can be obtained from the growth. In order to avoid degradation, the F4CGT single crystals are stored in a Ar-filled glovebox.
The microscopy images was acquired using a Bruker Edge Dimension AFM.

\noindent\textbf{Device fabrication.} 
F4CGT flakes were cleaved from bulk crystals onto polydimethylsiloxane (PDMS) by mechanical exfoliation and they were examined by an optical microscope to evaluate the thickness roughly. 
Then the atomically smooth flakes with desired thicknesses were transferred to a 285 nm SiO$_{2}$/Si substrate with pre-patterned electrodes and an $h$-BN capping layer was used to cover the sample for protection from H$_{2}$O and O$_{2}$. 
The Ti/Au (10/40 nm) electrodes was fabricated by electron beam lithography and metals were deposited using thermally evaporating method.
After transport measurements, the $h$-BN capping layer was removed and the thickness of sample was determined precisely by AFM. 
The whole fabrication process of device was carried out in an argon glove box with H$_{2}$O and O$_{2}$ contents less than 0.1 ppm to avoid degradation of the samples. 

\noindent\textbf{Magnetization and electrical transport measurements.}
Magnetization and electrical transport measurements were performed in a Quantum Design MPMS3 and PPMS. Both longitudinal and Hall electrical resistance were measured using a five-probe method on F4CGT Hall bar device with current flowing in the $ab$ plane. 
The Hall resistance was measured by sweeping the field from -5 T to +5 T and from +5 T to -5 T at various temperatures, and the total Hall resistance was determined by a standard symmetrization procedure to remove the contribution of magnetoresistance from the raw Hall voltage data due to voltage probe misalignment\cite{Ohno2}.  
The angular dependence of Hall resistance was measured using the PPMS Horizontal Rotator with the field rotating from $c$ axis to the current direction, which is always perpendicular to the direction of Hall voltage. 

\noindent\textbf{Magnetic characterization by MOKE.} 
Low temperature magnetic hysteresis loops of the exfoliated F4CGT flakes were measured by an MOKE system with magnetic field along the $ab$-plane. The MOKE system uses a 408 nm diode laser with a laser spot size $\sim$ 2 $\mu$m, and has Kerr rotation detection sensitivity $\textless$ 0.1 mrad.

\noindent\textbf{Theoretical calculations.}  
The first-principles electronic structure calculations on F4CGT were carried out by using the projector augmented wave method\cite{Blochl,Kresse1} as implemented in the VASP package\cite{Kresse2,Kresse3}. The generalized gradient approximation of Perdew-Burke-Ernzerhof\cite{Perdew} type was employed for the exchange-correlation functional. The kinetic energy cutoff of the plane-wave basis was set to be 500 eV. The $\Gamma$ centered 12$\times$12$\times$2 and 12$\times$12$\times$1 $k$-point meshes\cite{Monkhorst}  were adopted for the Brillouin zone sampling of the primitive cells of bulk and thin films, respectively. A vacuum layer larger than 2.2 nm was used for the thin film simulations to avoid the artificial interaction between periodic images. The Gaussian smearing technique with a width of 0.05 eV was utilized for the Fermi surface broadening. The lattice constants and the internal atomic coordinates were fully relaxed until the forces on atoms were smaller than 0.1 eV/nm. The DFT-D2 method\cite{Grimme,WuX} was used to include the vdW interactions for multilayers. When calculating the band structure and the magnetic anisotropy energy, the spin-orbit coupling effect was taken into consideration. The AHCs were calculated with the WannierTools software\cite{WuQ} based on the tight-binding Hamiltonians constructed from maximally localized Wannier functions obtained from the Wannier90 package\cite{Mostofi}.

\section{Results and discussion}

\begin{figure}
	\centerline{\includegraphics[scale=0.17]{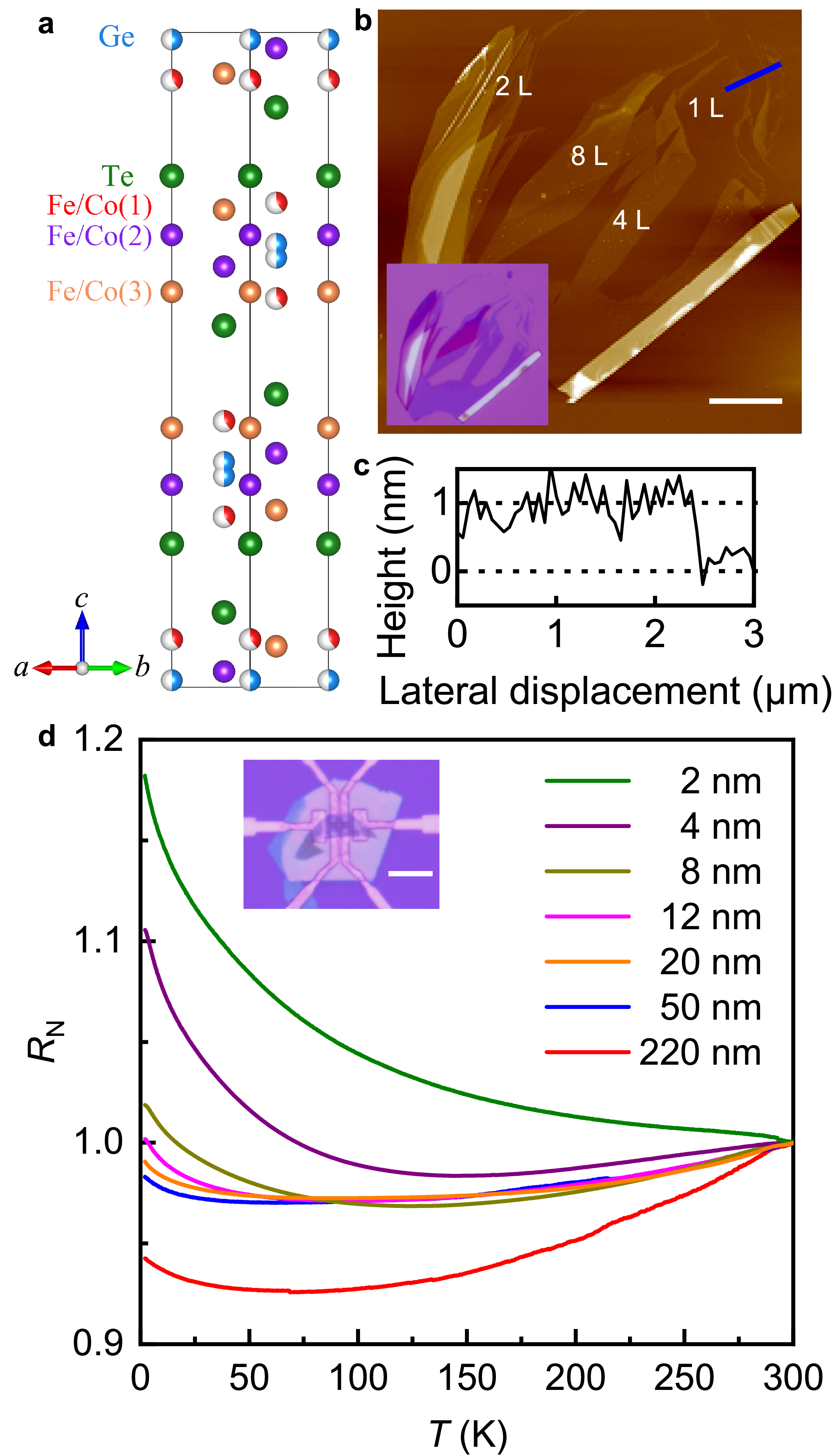}}
	\caption{\textbf{Crystal structure and characterization of F4CGT thin flakes.} \textbf{a}, Crystal structure of F4CGT. The blue and green balls represent Ge and Te atoms, respectively, and the red, purple and orange balls represent Fe/Co atoms at Fe(1), Fe(2) and Fe(3) sites, respectively. The Fe(1) and Ge positions are partially occupied (marked by the color difference).
		\textbf{b}, Atomic force microscope image of F4CGT thin flakes on a 285 nm SiO$_{2}$/Si substrate. The white scale bars represent 5 $\mu$m.  Inset: optical picture of F4CGT thin flakes.
		\textbf{c}, Cross-sectional profile of the F4CGT thin flake along the blue line in b.
		\textbf{d}, Temperature dependence of $R_{\rm N}(T)$ for F4CGT with various thicknesses. Inset: the optical image of a typical F4CGT device covered with a h-BN layer. The white scale bar represents 15 $\mu$m.}
\end{figure}

As shown in Fig. 1a, F4CGT has a layered structure with a rhombohedral $R\bar{3}m$ space group ($a=$ 0.40 nm and $c=$ 2.93 nm) and in one unit cell there are three Fe/Co-Ge-Te layers which stack in the ABC sequence along the $c$ axis via the weak vdW interaction\cite{Andrew,Tian,Andrew2}. Thus, the $t$ of single Fe/Co-Ge-Te layer is about 0.98 nm.
In a Fe/Co-Ge-Te layer, the Fe/Co-Ge slab is sandwiched by the Te atoms and because of inequivalent local environment of Fe atoms, there are three Fe sites.
The Fe(2) and Fe(3) positions are fully occupied, but the Fe(1) site has obvious vacancies\cite{May,Andrew}. Correspondingly, the Ge position splits into two sites to maintain an appropriate bond distance between Fe(1) and Ge\cite{May,Stahl,Andrew}. 
Importantly, when 20\% Co atoms are doped into the Fe position, they occupy primarily on the Fe(1) site, which may be related to the enhanced $T_{\rm C}$ and the change of easy direction of magnetization from the $c$ axis to the $ab$ plane\cite{Andrew2}. 
Fig. 1b illustrates the optical image of F4CGT flakes on a 285 nm SiO$_{2}$/Si substrate, which are mechanically exfoliated from a F4CGT single crystal with $T_{\rm C}\sim$ 335 K (Fig. S1 in the Supporting Information). F4CGT flakes with different $t$ can be resolved by the optical contrast. Combined with the morphology measurements using an atomic force microscope (AFM) (Fig. 1c), the relationship between optical contrast and $t$ can be established. As shown in Fig. 1d, a sharp step in the height profile for the thinnest F4CGT flake along the blue line on Fig. 1d can be observed and it is about 1 nm, which perfectly match the thickness of ML F4CGT. It clearly indicates that the atomically thin F4CGT can be obtained from mechanical exfoliation.

The inset of Fig. 1e displays a typical device of F4CGT with bottom Ti/Au electrodes and a top $h$-BN protection layer on the SiO$_{2}$/Si substrate. Using such devices, we measured the temperature dependence of normalized longitudinal resistance $R_{\rm N}(T)=R(T)/R($300 K$)$ for F4CGT with different $t$ (Fig. 1e). 
For the thick F4CGT sample ($t\sim$ 220 nm) the $R_{\rm N}(T)$ show a metallic behaviour with weak temperature dependence, which is consistent with that of F4CGT bulk crystal but distinctly different from the F5GT one\cite{Tian}.
For the latter, there is a kink around $T\sim$ 145 K, which is attributed to the magnetic ordering of Fe(1) site below this temperature\cite{May,Chen}. For F4CGT bulk crystal, such kink becomes diminished, implying the suppression of this transition by Co doping\cite{Tian}.
With reducing $t$, the metallic behaviour becomes weaker and there is a crossover from metallic behaviour to insulating one that shifts to higher temperature gradually. 
When $t\leq$ 4 nm, the $R_{\rm N}(T)$ increases rapidly with decreasing temperature. While for the BL F4CGT ($t=$ 2 nm), the $R_{\rm N}(T)$ exhibits an insulating behaviour in the whole measuring temperature region (2 - 300 K). Such increase of insulating behaviour with decreasing $t$ is very similar to the case in F3GT but different from that of F5GT which still shows a metallic behaviour when the $t$ is down to 6.8 nm\cite{Deng,Tan}.

\begin{figure}
	\centerline{\includegraphics[width=\linewidth]{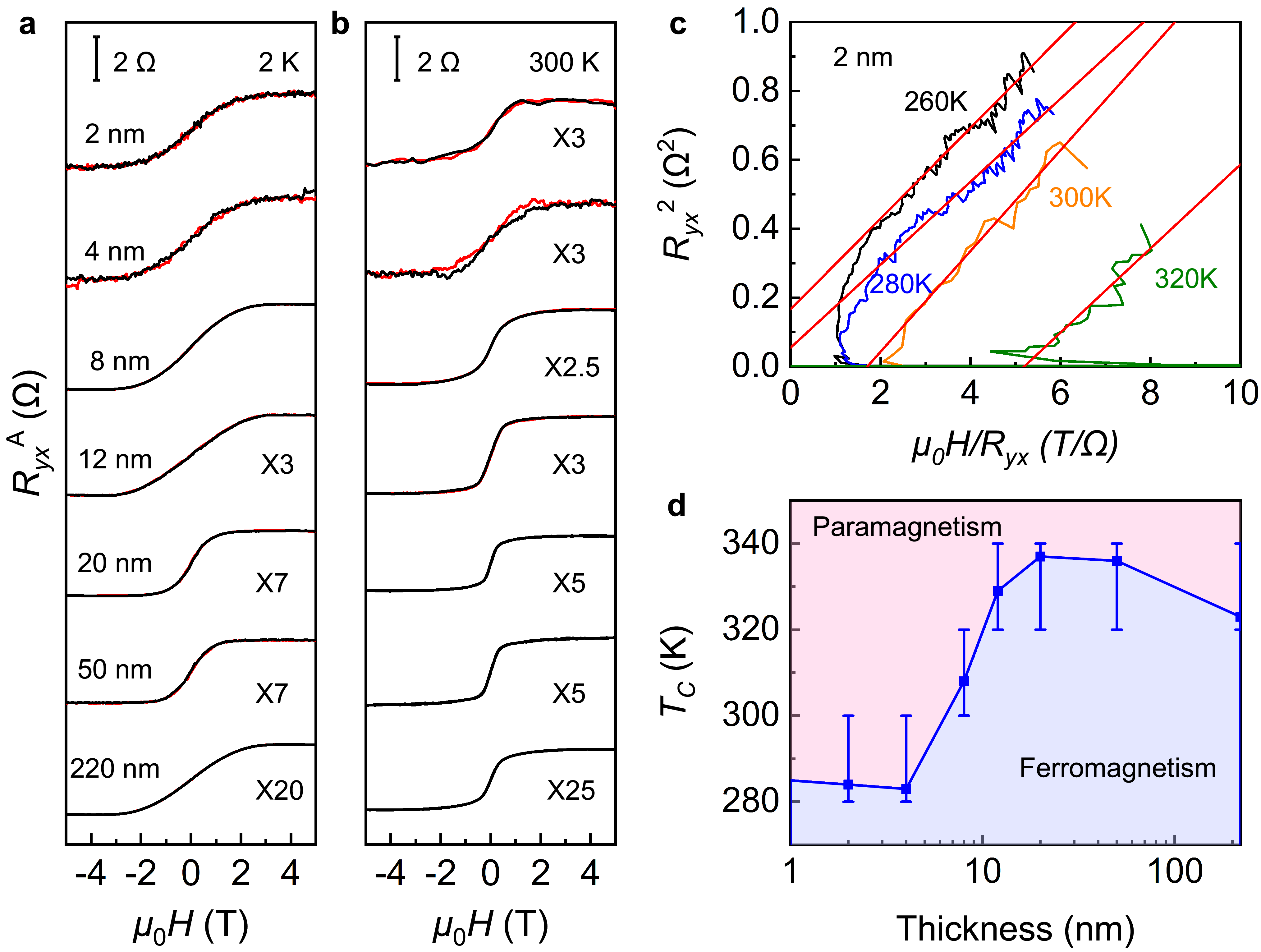}}
	\caption{\textbf{Nearly-room-temperature ferromagnetism in F4CGT thin flakes.} \textbf{a} and \textbf{b}, Field dependence of $R_{yx}^{\rm A}(\mu_{0}H)$ measured at 2 K and 300 K for the F4CGT thin flakes with various $t$, respectively. The red and black lines represent the $R_{yx}^{\rm A}(\mu_{0}H)$ measured when decreasing and increasing fields. 
		\textbf{c}, Arrott plots of a 2 nm device near 300 K. The red lines are linear fits at high-field region.
		\textbf{d}, Phase diagram of F4CGT as functions of $T$ and $t$. The $T_{\rm C}$ values (blue squares) are derived by linear interpolations of two adjacent Arrott plots. Vertical error bars represent the temperatures corresponding to two Arrott plots used to determine $T_{\rm C}$. Pink and blue regions mark the FM and PM states, respectively.
	}
\end{figure}

In order to probe the dependence of long-range ferromagnetism in F4CGT on $t$, Hall resistance $R_{yx}$ were measured for F4CGT Hall bar devices with various $t$ and all measurements were performed with $H\Vert c$.
As shown in Fig. 2a, after subtracting the ordinary Hall resistance $R_{yx}^{\rm O}(\mu_{0}H)$ obtained from the linear fit of high-field $R_{yx}(\mu_{0}H)$ data (Fig. S2 in the Supporting Information), the anomalous Hall resistance $R_{yx}^{\rm A}(\mu_{0}H)$ curves at 2 K for all of F4CGT samples show a similar S-shape and the values of $R_{yx}^{\rm A}(\mu_{0}H)$ saturate at high-field region. 
In addition,the Arrott plots of $R_{yx}^{2}$ vs. $\mu_{0}H/R_{yx}$ at 2 K  are shown in Fig. S3 in the Supporting Information. For the Arrott plot, the positive and negative value of the $y$-axial intercept indicates the FM and paramagnetic (PM) states, respectively, and the temperature corresponding to zero-value intercept is $T_{\rm C}$\cite{Arrott,Ohno1,Chiba}. It can be seen that the $y$-axial intercepts obtained from the high-filed linear fits are positive for all of F4CGT samples with different $t$. These results clearly indicate that at low temperature F4CGT is still in ferromagnetic state even for the BL sample.
In contrast to F3GT and F5GT\cite{Deng,Tan}, the $R_{yx}^{\rm A}(\mu_{0}H)$ curves of F4CGT samples do not show obvious hysteresis behaviours, similar to the isothermal magnetization curve $M(\mu_{0}H)$ for $H\Vert c$ in bulk crystal (Fig. S1 in the Supporting Information). Moreover, such soft magnetism can be partially ascribed to the field direction parallel along the hard axis of magnetization of F4CGT\cite{Tian,Andrew2}. 
On the other hand, although thermal fluctuation weakens the FM state at 300 K, leading to the reduction of saturation field $\mu_{0}H_{\rm s}$, the saturation behaviours of $R_{yx}^{A}(\mu_{0}H)$ with nonzero saturation values $R_{yx}^{\rm{A,s}}$ persist at 300 K for all of samples with various $t$, even for the BL F4CGT (Fig. 2b). These results strongly suggest that a nearly-room-temperature ferromagnetism could be realized in ultra thin F4CGT crystals without gating or exerting pressure. 

Fig. 2c and Fig. S4 in the Supporting Information show the Arrott plots at high-temperature region for the samples with different $t$ in order to determine $T_{\rm C}$ accurately.
For the BL sample (Fig. 2c), the positive $y$-axial intercept at 280 K suggests that the FM state persists at $T>$ 280 K and the $T_{\rm C}$ determined from the linear interpolation of two adjacent curves in the Arrott plot is about 284 K.
In addition, we have also carried out the measurement of planar Hall effect (PHE) of bilayer F4CGT (Fig. S5  in the Supporting Information).
The planar Hall resistance (PHR) $R_{yx}^{\rm PHE}(\phi)$ as a function of the angle $\phi$ between the in-plane magnetic field ($\mu_{0}H$ = 5 T) and the direction of electrical current $I$ also show a two-fold sin(2$\phi$)-like oscillation at low temperatures, which is consistent with the angular dependence of PHR for a single domain ferromagnet with in-plane magnetization [$R_{yx}^{\rm PHE}(\phi)$ = ($R_\Vert - R_\perp$)sin(2$\phi$)/2], where $R_\Vert$ is longitudinal resistance for $H\Vert I$ and $R_\perp$ is transverse resistance for $H\perp I$.
Moreover, it can be seen that when temperature is larger than 285 K, the oscillation feature disappears. This is a proof that the ferromagnetic-paramagnetic phase transition happens at about 285 K in 2nm-thick F4CGT, in agreement with the results obtained from the analysis of Arrott plots.
Moreover, 	the measurements of magneto-optic Kerr effect (MOKE) also confirms the existence of a nearly-room-temperature $ab$-plane ferromagnetism of F4CGT thin flakes (Fig. S6 in the Supporting Information). For the F4CGT samples with $t\sim$ 10 and 15 nm, the hysteresis loop and residual Kerr rotation ($\theta_{\rm K}$) appear when $T\leq$ 280 and 310 K, respectively, which are hallmarks of FM ordering. The $\mu_{0}H_{\rm s}$ obtained from the MOKE measurements for $H\Vert ab$ are also smaller than those extracted from the Hall measurements for $H\Vert c$ (Fig. S7 in the Supporting Information), consistent with the in-plane magnetic anisotropy of F4CGT thin flakes.
Fig. 2d shows the phase diagram of F4CGT samples with different $t$ and the $T_{\rm C}$ values are estimated from the Arrott plots.
For the samples with $t\geq$ 12 nm, the $T_{\rm C}$s fall in the range of 320 K - 340 K, which are close to the bulk value ($\sim$ 335 K). For the thinner F4CGT samples, the $T_{\rm C}$ starts to decrease and reaches about 284 K in a BL F4CGT. 
This can be explained by the dimensional effect because the density of states per spin for the magnon modes near the excitation gap increases as the number of layers decreases and the magnon excitations can destroy the long-range magnetic order at a lower $T_{\rm C}$\cite{Gong1,Deng}.
Moreover, unlike the completely suppressed $T_{\rm C}$ to zero in the 2D isotropic Heisenberg system because of thermal fluctuations of the long-wavelength gapless magnon modes, the finite $T_{\rm C}$ in atomically thin F4CGT suggests that there is a magnetocrystalline anisotropy in this system, which can give rise to an energy gap in the magnon dispersion and establish a ferromagnetic order in 2D at finite temperature\cite{Gong1,Deng}.
But the limited range of $t$ in present work impedes extracting the critical exponent of transition and the critical spin-spin coupling range.

\begin{table}
	\caption{The calculated relative energies of the FM states with respect to the corresponding nonmagnetic (NM) states and the magnetic anisotropy energies (MAE) for F4CGT bulk, BL, and ML. Bulk structures are established by the ABC stacking. The “disorder” means that the Fe(1) sites are randomly distributed on the sides of each F4CGT layer. The “Fe(1)a only” means that the Fe(1) sites are on the same sides of each F4CGT layer and Co atoms only occupy the Fe(1)a sites. The BLs are constructed with different Co occupations of Fe(1) subsites and different stacking sequences. The energies are averaged to the ones per formula unit (f.u.).}
	\centering
	\begin{tabular}{ccc}
		\hline\hline      
		Energy/f.u.           & $\Delta E$ ($E_{\rm FM}-E_{\rm NM}$)  & MAE ($E_{M_z}-E_{M_x}$)  \\ 
		\hline
		Bulk (disorder)       & -1.240 eV                     & 0.503 meV               \\ 
		Bulk (Fe(1)a only)    & -1.344 eV                     & -0.076 meV               \\ 
		Bilayer (AA-stacking) & -1.379 eV                     & 0.070 meV                \\ 
		Bilayer (AB-stacking) & -1.352 eV                     & 0.454 meV                \\ 
		Monolayer             & -1.524 eV                     & 0.756 meV                \\ 
		\hline\hline   
	\end{tabular}
\end{table}

To obtain more insight into the ferromagnetism in F4CGT thin flakes, we carried out the first-principles electronic structure calculations for bulk and few-layer F4CGT. 
According to the previous studies\cite{Tian, Andrew2}, the magnetism of Co doped F5GT is sensitive to the layer stacking and the Co substitution. Therefore, we have studied the ML, BL, and bulk F4CGT with the substituted Co atoms at different Fe sites and with different layer stackings. 
In the ML F4CGT, we found that the most stable substitution site for Co is the Fe(1) site, whose energy is 0.117 eV lower than the second lowest one, i.e., the Fe(2) site. Meanwhile, the FM state is energetically much lower than the nonmagnetic (NM) state, and the easy direction of magnetization is in  the $ab$ plane (Table I). 
These are consistent with previous theoretical results\cite{Andrew2,ChenH}. 
For the BL F4CGT, we found that the favorable substitution site is the same as the ML case, namely the Fe(1) site. Here the interlayer stacking of BL F4CGT could be AA- or AB-stacking. No matter in which stacking, the energy of the FM state is always much lower than that of the NM state (Table I) and slightly (several meV/f.u.) lower than that of the interlayer antiferromagnetic state (Fig. S8 and Table S1  in the Supporting Information). 
This indicates that F4CGT has a strong intralayer FM coupling but a weak interlayer interaction due to the vdW gap. The magnetic anisotropy energy of BL F4CGT depends on the layer stacking, but the easy direction of magnetization keeps in the $ab$ plane (Table I). 
As to bulk F4CGT, we adopted the experimental ABC-stacking sequence. The magnetic ground state of bulk F4CGT is also the FM state (Table I). Depending on the distribution of Fe(1) sites in each layer, the calculated easy direction of magnetization could be either the $ab$ plane (disorder) or the $c$ axis (Fe(1)a only). In the real material, the Fe(1) atoms distribute randomly in each FCGT layer to maximize the entropy\cite{May,ZhengQ}, hence the easy direction of magnetization is in the $ab$ plane.
Overall, our calculations show that the bulk and few layer F4CGT adopt the FM ground state with the magnetization direction in the $ab$ plane, which are in good agreement with the experimental observations.

\begin{figure}
	\centerline{\includegraphics[width=\linewidth]{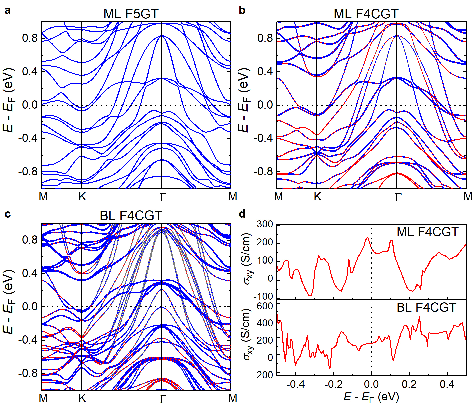}}
	\caption{\textbf{Theoretical calculation results of F5GT and F4CGT.} \textbf{a-c}, Band structures calculated with the spin-orbit coupling for the ground states (i.e., in-plane FM states) of ML F5GT, ML F4CGT and AA-stacking BL F4CGT with Co occupying at Fe(1) site, respectively. The points in red and blue in b and c represent the contributions from Co and Fe atoms, respectively.
		\textbf{d}, The in-plane anomalous Hall conductivity $\sigma_{xy}$ of ML and AA-stacking BL F4CGT as a function of $E-E_{\rm F}$. Here the magnetization is aligned to the $c$-axis and $E_{\rm F}$ is the Fermi level.}
\end{figure}

We next make a comparison between the ML F5GT and F4CGT. On the one hand, the calculated FM state of ML F4CGT (F5GT) is 1.524 (1.275) eV/f.u. lower than that of the NM state, indicating that Co substitution can enhance the ferromagnetism.
On the other hand, the calculated local moments for Fe(1) atom in F5GT and Co atom at Fe(1) site in F4CGT are 0.04 $\mu_{\rm B}$ and 0.34 $\mu_{\rm B}$, respectively, while their total moments per unit cell are 8.30 $\mu_{\rm B}$ and 8.76 $\mu_{\rm B}$. The total magnetic moments calculated here are similar to the value of 2 $\mu_{\rm B}$/Fe observed in F5GT, i.e., about 9 $\mu_{\rm B}$/f.u.\cite{May}. Due to the more stable FM state and the larger local moments, F4CGT films can own a higher $T_{\rm C}$ than F5GT\cite{Tian, Andrew2}. 
In addition, the calculated band structures show that there is a small upshift of the $E_{\rm F}$ for F4CGT when compared with F5GT since the Co dopants have introduced more valence electrons (Figs. 3a and b). The Co dopants also modify the dispersion of some bands near the $E_{\rm F}$ (Figs. 3b, c and Fig. S9 in the Supporting Information), which could influence the itinerant carriers and magnetism of F4CGT.
For the BL and bulk F4CGT (Fig. 3c and Fig. S10 in the Supporting Information), the stacking of layers introduces multiple bands compared with the ML one, but no prominent changes take place in the band structure due to the weak vdW-type interlayer interaction. 

\begin{figure}
	\centerline{\includegraphics[width=\linewidth]{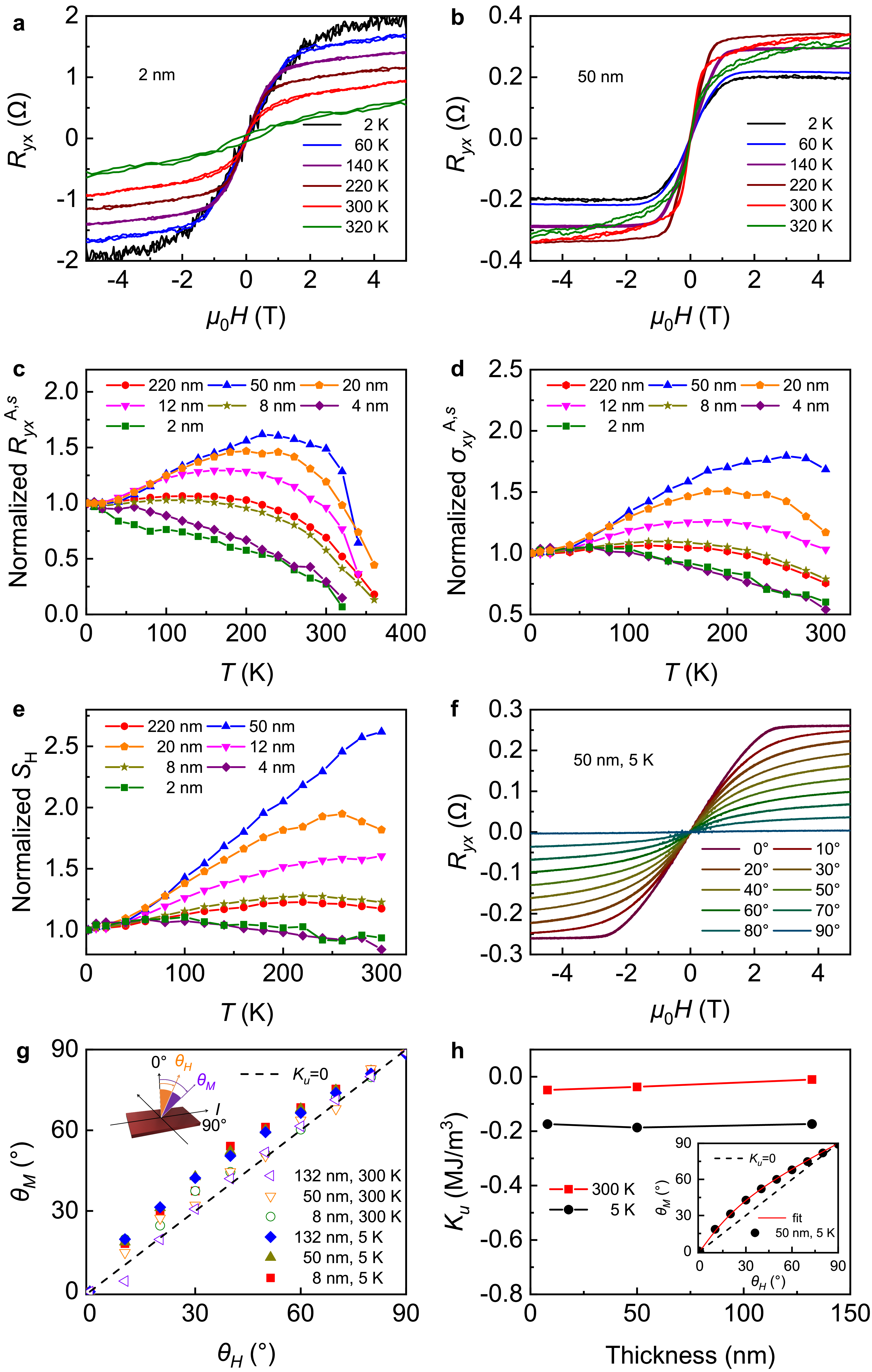}}
	\caption{\textbf{Evolution of anomalous Hall effect with $T$ and $t$.} \textbf{a} and \textbf{b}, Field dependence of $R_{yx}(\mu_{0}H)$ at different temperatures for the samples with $t$ = 2 nm and 50 nm, respectively. \textbf{c-e}, Temperature dependence of normalized $R_{yx}^{\rm{A,s}}$, $\sigma_{yx}^{\rm{A,s}}(T)$ and $S_{\rm H}(T)$ for the F4CGT samples with various $t$. These data are normalized by their values at $T$ = 2 K.
		\textbf{f},  Angular dependence of the $R_{yx}(\mu_{0}H)$ curves at $T$ = 5 K for sample with $t$ = 50 nm.
		\textbf{g}, The relationship of $\theta_{M}$ and $\theta_{H}$ at 5 K and 300 K for F4CGT samples with different $t$.  The black dashed line marks $\theta_{M}$ = $\theta_{H}$ that corresponds to $K_{u}$ = 0. Inset shows the definition of  $\theta_{M}$ and $\theta_{H}$.
		\textbf{h}, Temperature and thickness dependence of fitted $K_{u}$ using the Stoner-Wohlfarth model. Inset exhibits the fit of  $\theta_{M}(\theta_{H})$ curve at 5 K for 50 nm thick sample and the black dashed line is same as that in inset of \textbf{g}.}
\end{figure}

After confirming the existence of nearly-room-temperature ferromagnetism in atomically thin F4CGT, we will discuss anther feature of F4CGT that the $R_{yx}(\mu_{0}H)$ curves of samples with different $t$ exhibit different trends with the change of temperature (Figs. 4a, b and Fig. S7 in the Supporting Information).
As shown in Fig. 4a, the low-field $R_{yx}(\mu_{0}H)$ curves  at different temperatures for the BL sample almost fall on the same line and the high-field $R_{yx}(\mu_{0}H)$ decreases gradually with increasing $T$. Meanwhile, both $\mu_{0}H_{\rm s}$ and $R_{yx}^{\rm{A,s}}$ decrease monotonically. 
In contrast, for the 50 nm sample (Fig. 4b), the high-field $R_{yx}(\mu_{0}H)$ curves at low temperatures are almost flat and they increase with field at high temperatures.
Moreover, although the $\mu_{0}H_{s}$ decreases monotonically with increasing $T$, the $R_{yx}^{\rm{A,s}}$ exhibit a non-monotonic $T$ dependence, i.e., it first increases  and then decreases with increasing $T$.
Fig. 4c show the temperature dependence of $R_{yx}^{\rm{A,s}}(T)$ which are normalized by their values at $T=$ 2 K for the samples with different $t$. 
For the thickest sample ($t=$ 220 nm), the normalized $R_{yx}^{\rm{A,s}}(T)$ is almost unchanged with $T$ warming up to about 200 K and then starts to decrease gradually when $T$ increases further.  
But for the 50 nm thick sample, the temperature-dependent normalized $R_{yx}^{\rm{A,s}}(T)$ changes dramatically and it exhibits a dome-shaped behaviour.
When further decreasing $t$, the peak position of $R_{yx}^{\rm{A,s}}(T)$ curve shifts to lower temperatures gradually, and finally such dome-shaped behaviour disappears for the samples with $t=$ 4 nm and 2 nm.
Similar non-monotonic behaviours of $R_{yx}^{\rm{A,s}}(T)$ shown in samples with medium $t$ have been observed in F4GT and F5GT thin flakes, in which the peak position of $R_{yx}^{\rm{A,s}}(T)$ is closely related to the characteristic temperature of spin-reorientation process from $ab$ plane to $c$ axis\cite{Seo,LiZY}. In contrast, for the $c$-axial ferromagnet F3GT thin flakes with various $t$, such behaviour is absent and the $R_{yx}^{\rm{A,s}}(T)$ shows a monotonic decrease when $T$ increases\cite{Deng}.

For ferromagnet with intrinsic anomalous Hall effect (AHE) related to the Berry curvature at reciprocal space, it has $ \rho_{yx}^{\rm A}(T) = S_{H}(T)\rho_{xx}^{2}(T)M(T)$, where $ \rho_{yx}^{\rm A} $ is anomalous Hall resistivity, $S_{H}$ is the anomalous Hall factor and $\rho_{xx}$ is the longitudinal resistivity\cite{Nagaosa}. 
Moreover, because the AHC $\sigma_{xy}^{\rm A}(T)\approx \rho_{yx}^{A}(T)/\rho_{xx}^{2}(T)$ when $\rho_{xx}(T)\gg \rho_{yx}^{\rm A}(T)$, it has $\sigma_{xy}^{\rm A}(T)=S_{\rm H}(T)M(T)$.
Thus, the temperature dependence of $ \rho_{yx}^{\rm A}(T)$ originates from the temperature-dependent  $S_{\rm H}(T)$, $\rho_{xx}(T)$ and/or $M(T)$. 
First, the temperature-dependent $\rho_{xx}(T)$ can lead to the non-monotonic  $\rho_{yx}^{\rm{A}}$, which has been observed in Co$_{3}$Sn$_{2}$S$_{2}$ and Yb$_{14}$MnSb$_{11}$\cite{WangQ,Brain}. In order to eliminate the influence of $R_{xx}(T)$ on $ R_{yx}^{\rm A}(T)$, we plot the normalized saturation Hall conductivity $\sigma_{xy}^{\rm {A},s}(T)=[R_{yx}^{\rm{A,s}}(T)/R_{xx}^{2}(T)]/[R_{yx}^{\rm{A,s}}(2$ K$)/R_{xx}^{2}(2$ K$)]$ as a function of $T$ (Fig. 4d).
The curves show similar behaviours to those $R_{yx}^{\rm {A,s}}(T)$, i.e., there is still an obvious non-monotonic temperature dependence of $\sigma_{xy}^{\rm{A,s}}(T)$ when $t$ is between 8 and 220 nm.
Second, in many of itinerant ferromagnets, the dependence of $\sigma_{xy}^{\rm A}(T)$ on $T$ derives entirely from that of $M(T)$, in other words,  the $\sigma_{xy}^{\rm A}(T)$ is linearly proportional to $M(T)$ with constant $S_{\rm H}$\cite{WangQ,Lee,ZengC}.
Fig. 4e shows the dependence of  normalized $S_{\rm H}=\sigma_{xy}^{\rm {A},s}(T)/M(T)$ on $T$, where the $M(T)$ is the value of bulk crystal measured at $\mu_{0}H=$ 5 T. 
It can be seen that for the F4CGT samples with $t\leq$ 8 nm or $t\geq$ 220 nm, the $S_{\rm H}$ exhibits a weak temperature dependence and for other samples it changes significantly with $T$. 
Here is a natural question whether the magnetic anisotropy of F4CGT varies with $t$. In other words,  if the spin-reorientation process appears in the samples with $t$ between 8 and 220 nm, there could be a remarkable change of $S_{\rm H}$, like the case in F4GT\cite{Seo}. 
To reveal the evolution of magnetic anisotropy of F4CGT with $T$ and $t$, the angle-dependent $R_{yx}(\mu_{0}H)$ were measured and Fig. 4f shows typical results at 5 K for the sample with $t=$ 50 nm. 
For all of angles, the $R_{yx}(\mu_{0}H)$ curves display a S-shape with negligible hysteresis loops and the $R_{yx}($5 T$)$ decreases with the increase of $\theta_{H}$ (the angle between external field direction and $c$ axis) (inset of Fig. 4g). Because  the $R_{yx}$ is proportional only to the out-of-plane component of magnetization, there is a relationship between $R_{yx}$, $\theta_{H}$ and $\theta_{M}$, $R_{yx}(\theta_{H})=R_{yx}(\theta_{H}=0^{\circ})\cos\theta_{M}$, where $\theta_{M}$ represents the angle between magnetization direction and $c$ axis) (inset of Fig. 4g). When using the values of $R_{yx}($5 T$)$ at various $\theta_{H}$s, the values of $\theta_{M}$ are calculated using the formula $\theta_{M} = \arccos(R_{yx}(\theta_{H})/R_{yx}(\theta_{H} = 0^{\circ}))$.
As shown in Fig. 4g, the $\theta_{M}$ are always larger than $\theta_{H}$ for all of samples at both 5 K and 300 K, clearly indicating that the magnetization always tends to point to the $ab$ plane regardless of the direction of external magnetic field.
These results are also consistent with the MOKE results (Fig. S6 in the Supporting Information).
Furthermore, the magnetic anisotropy energy density $K_{u}$ can be derived by fitting the $\theta_{M}-\theta_{H}$ data using the Stoner-Wohlfarth model\cite{Stoner} with $\partial E/\partial\theta_{M}=2K_{u}\sin\theta_{M}\cos\theta_{M}-\mu_{0}HM_{s}\sin(\theta_{H}-\theta_{M})=0$, where $E$ is the total energy of a magnetic system and $M_{s}$ is the saturation moment at 5 T. 
As shown in Fig. 4h, the fitted values of $K_{u}$ at 5 K and 300 K are negative and insensitive to $t$. The values of $K_{u}$ at 300 K become smaller than those at 5 K because $|K_{u}|\propto M_{s}^n$ with $n>$ 1 and the $M_{s}$ decreases with the increase of $T$\cite{Okamoto}. 
In addition, the $|K_{u}|$ at 5 K of F4CGT is relatively small when compared to F3GT ($K_{u}\sim$ 0.51 MJ/m$^{3}$) and F4GT ($K_{u}\sim$ 0.39 MJ/m$^{3}$) \cite{Deng,Seo}, which reflecting the soft magnetism of F4CGT as well.
In a word, same as bulk crystal, atomically thin F4CGT is still an easy-plane ferromagnet without spin-reorientation process when decreasing $T$.Above results clearly indicate that the strong $T$ dependence of $S_{\rm H}(T)$, i.e., the nonlinear dependence of $\sigma_{xy}^{\rm {A},s}(T)$ on $M(T)$ for the samples with medium $t$, do not originate from the change of magnetic anisotropy.

To understand the experimentally observed AHE of F4CGT, we calculated the $\sigma_{xy}$ as a function of $E-E_{\rm F}$ for the ML and AA-stacking BL F4CGT (Fig. 3d), which exhibit quite different behaviours around the $E_{\rm F}$.
If assuming the $E_{\rm F}$ moves towards the higher energy along with the increasing temperature, for which the room temperature (about 300 K) corresponds to an upward energy shift of about 26 meV, the AHC of ML F4CGT will decrease with temperature while that of BL one will increase.
As for the BL F4CGT, the energy of AB stacking structure is only 36 meV/f.u. higher than the AA-stacking one, we also calculated the AHC of AB stacking BL F4CGT, whose behaviour is similar to that of AA-stacking bilayer F4CGT, except for a smaller slope of AHC with respect to the energy around the $E_{\rm F}$ than that of the AA-stacking one (Fig. S10 in the Supporting Information). 
Hence, these results explain the evolution of AHC with $T$ and $t$ observed in experiment qualitatively.
The similar peculiar behaviour of nonlinear dependence of $\sigma_{xy}^{\rm {A},s}(T)$ on $M(T)$ (or non-monotonic dependence on $T$) shown in F4CGT has also been observed in Sr$_{1-x}$Ca$_{x}$RuO$_{3}$ previously\cite{FangZ,Mathieu}. It is explained by the existence of band crossing point in the band structure which act as a monopole for the gauge field representing the Berry curvature. The susceptive feature of Berry curvature distribution causes the complex change of $\sigma_{xy}^{\rm {A},s}(T)$ with $M(T)$ because the $E_{\rm F}$ crosses the monopole energy as $T$ changes\cite{FangZ,Mathieu}. 
For F4CGT, the position of $E_{\rm F}$ changes with $T$ and the band structure with several band crossing points near $E_{\rm F}$ also evolves with $t$ subtly (Figs. 3b and c), thus the drastically changed Berry curvature near $E_{\rm F}$ could lead to the unusual temperature- and thickness-dependent intrinsic AHE.

\section{Conclusions}
In summary, present work reveals that a nearly-room-temperature ferromagnetism can be realized in F4CGT at 2D limit. 
Interestingly, the intrinsic AHC in F4CGT with $t$ between 8 nm and 220 nm shows a non-monotonic dependence on $T$, but this behaviour can not be explained by the spin-reorientation process because the easy-plane ($ab$-plane) ferromagnetism in F4CGT is robust and does not change with $T$ and $t$.Combined with theoretical calculations, we propose that the thickness-dependent electronic structure and the temperature-dependent shift of $E_{\rm F}$ may be the reasons of the non-monotonic behaviour of $\sigma_{xy}^{\rm {A},s}(T)$ curves.
The discovery of nearly-room-temperature ferromagnetism and tunable AHC in F4CGT not only expands the family of 2D room-temperature ferromagnetic materials which are important to the area of spintronics, but also provides a novel platform to study the unique features of intrinsic AHE in 2D vdW magnetic materials.

\section{Author Contributions}

H.C.L. provided strategy and advice for the research; S.H.Y. Q.W.Y. and X.Y.C. grew the single crystal; S.H.Y., Y.F., S.J.T. and F.Y.M. prepared the devices and performed the measurements of physical properties with the assistance of L.W. and S.S.C.; S.H.Y performed the fundamental data analysis with the assistance of S.G.W., X.Z. and H.C.L.;  K.H.S. J.W.C. and H. R. performed the measurements of  MOKE. H.H.H., N.N.Z. and K.L. carried out the theoretical calculation; S.H.Y., H.H.H and H.C.L. wrote the manuscript based on discussion with all the authors.

\section{Conflicts of interest}
There are no conflicts to declare.

\section{Acknowledgements}
This work was supported by National Key R\&D Program of China (Grants Nos. 2018YFE0202600 and 2022YFA1403800), Beijing Natural Science Foundation (Grant No. Z200005), National Natural Science Foundation of China (Grants Nos. 12274459 and 12174443), the Fundamental Research Funds for the Central Universities and Research Funds of Renmin University of China (RUC) (Grants Nos. 18XNLG14, 19XNLG17 and 21XNLG26), the Outstanding Innovative Talents Cultivation Funded Programs 2022 of Renmin University of China, and Beijing National Laboratory for Condensed Matter Physics. Computational resources were provided by the Physical Laboratory of High-Performance Computing at Renmin University of China and Beijing Super Cloud Computing Center. H. R. acknowledges support from the National Research Foundation of Korea (NRF) grant funded by the Korea government (MSIT) (No. 2021R1A2C2014179, 2020R1A5A1016518).

$^{\dag}$ These authors contributed equally to this work.

$^{\ast}$ Corresponding authors: X.Z. (zhangxiaobupt@bupt.edu.cn); K.L. (kliu@ruc.edu.cn); H.C.L. (hlei@ruc.edu.cn).

\end{document}